\documentclass[letterpaper]{raa}           
\usepackage{graphicx,times}
\usepackage{natbib}
\usepackage{amssymb,amsmath}
\usepackage{soul}
\usepackage{xcolor}
\bibpunct{(}{)}{;}{a}{}{,}
\usepackage[pagebackref=true]{hyperref}

\newcommand{\rd}{r_d}
\newcommand{\DM}{D_{\!M}}
\newcommand{\DHub}{D_{\!H}}

\newcommand{\rev}[1]{\textcolor{black}{#1}}
\newcommand{\plotorplaceholder}[2]{%
  \IfFileExists{#1}{\includegraphics[width=#2]{#1}}{%
    \fbox{\parbox[c][0.25\textheight][c]{#2}{\centering Placeholder for figure\\[4pt]\ttfamily\detokenize{#1}}}%
  }%
}

\begin{document}

\title{Low-redshift-agnostic BAO Constraints on Binned Dark-energy Density Evolution from DESI DR1 and DR2}

\volnopage{ {\bf 20XX} Vol.\ {\bf XX} No. {\bf XX}, 000--000}
\setcounter{page}{1}

\author{Qian-Mo Liu
   \inst{1}
   \and Gong-Bo Zhao
   \inst{2}
}

\institute{
Experimental High School Attached to Beijing Normal University, No. 14 Erlong Road, Xicheng District, Beijing 100032, P.R. China; liam071207@gmail.com\\
\and
National Astronomical Observatories, Chinese Academy of Sciences, Beijing 100101, P.R. China; gbzhao@nao.cas.cn\\
\vs \no
{\small Received 20XX Month Day; accepted 20XX Month Day}
}

\abstract{
We present a low-redshift-agnostic compression of anisotropic baryon acoustic oscillation (BAO) distances to constrain the normalized dark-energy density evolution, $X(z)\equiv \rho_{\rm DE}(z)/\rho_{\rm DE}(0)$, above the lowest BAO redshift node \rev{$z_1$}. Standard BAO summaries include the transverse comoving distance $\DM(z)/\rd$, which depends on the integral of $H^{-1}(z)$ from $z=0$ to $z$ and therefore mixes the expansion history at $z<z_1$ with the higher-redshift signal. We instead replace the set $\{\DM(z_i)/\rd\}$ by adjacent increments $\{\Delta\DM(z_i,z_{i+1})/\rd\}$ while retaining the radial distances $\{\DHub(z_i)/\rd\}$. The mapping is linear, so the covariance propagates exactly. This compression intentionally removes one absolute transverse-distance mode, namely the additive contribution to $\DM/\rd$ below the first BAO node, and preserves the remaining information relevant to reconstructing the expansion history above $z_1$. Applied to DESI DR1 and DR2 anisotropic BAO measurements, the method yields almost uncorrelated constraints on piecewise-constant interval parameters $X_j$. In this sense, the compressed likelihood provides a conservative band-power-like estimate of dark-energy evolution: each interval is constrained mainly by BAO information from its own redshift range, while one nonlocal transverse mode and stronger global assumptions are deliberately projected out or marginalized over. Because our baseline analysis also marginalizes over bin-local matter-density and distance-scale parameters with broad external priors, the resulting $X_j$ constraints should be interpreted as a low-redshift-agnostic BAO baseline rather than as a fully prior-free reconstruction. All bins are consistent with $X=1$ within current uncertainties.
}

\authorrunning{Q.-M. Liu \& G.-B. Zhao}
\titlerunning{Low-redshift-agnostic BAO constraints on binned $X(z)$}

\maketitle

\section{Introduction}
\label{sec:intro}

The physical origin of cosmic acceleration remains one of the central open questions in modern cosmology. In the standard $\Lambda$CDM model the late-time expansion is driven by a cosmological constant, but a wide class of alternatives predict departures from a constant dark-energy density. A natural strategy is therefore to reconstruct the expansion history as flexibly as possible and to identify which redshift modes are actually constrained by the data \citep[e.g.,][]{HutererStarkman2003,WangTegmark2004,HutererCooray2005}. Recent DESI DR2 dark-energy studies have pushed this program further with both parametric and non-parametric reconstructions combined with supernova and CMB information \citep{DESI_DR2_ExtendedDE,DESI_DR2_DynamicalDE}, providing an important broader context for the deliberately conservative BAO-based baseline developed here.

Rather than adopting a specific functional form for $w(z)$, we focus on the normalized dark-energy density evolution,
\begin{equation}
X(z) \equiv \frac{\rho_{\rm DE}(z)}{\rho_{\rm DE}(0)},
\end{equation}
and describe it in redshift bins. Such reconstructions are intrinsically challenging because distance measurements are nonlocal integrals of the expansion rate. As a result, naive binned reconstructions often develop strong bin-to-bin correlations even when the underlying data appear local in redshift.

Baryon acoustic oscillations (BAO) provide a particularly robust geometric probe. The sound horizon at baryon drag, $\rd$, acts as a standard ruler calibrated by early-universe physics \citep{Eisenstein1998,Eisenstein2005}. Anisotropic BAO analyses constrain both the transverse comoving distance $\DM(z)/\rd$ and the radial distance $\DHub(z)/\rd$. The standard transverse-distance summary, however, depends on the integral of $H^{-1}(z)$ from $0$ to $z$, so every $\DM(z)/\rd$ point inherits an additive contribution from the poorly isolated low-redshift interval below the first BAO node.

The goal of this paper is to construct a transparent BAO-based baseline that is agnostic to that low-redshift contribution. Our key step is to replace the set of transverse distances at BAO nodes by adjacent increments,
\begin{equation}
\Delta\DM(z_i,z_{i+1}) \equiv \DM(z_{i+1}) - \DM(z_i),
\end{equation}
while retaining the measured $\DHub(z_i)/\rd$ nodes. This linear transformation propagates the covariance exactly. It does not preserve the full original BAO summary vector: instead, it intentionally projects out one absolute transverse-distance mode, namely the additive contribution to $\DM/\rd$ below the first BAO node. The remaining compressed data vector is the information relevant to reconstructing the expansion history above $z_1$.

We combine this low-redshift-free compression with a piecewise-constant description of $X(z)$ matched to the BAO redshift nodes. In our baseline implementation, each redshift interval is also assigned its own matter-density and distance-scale nuisance parameters. This choice is useful for localizing the inference, but it means that the inferred $X_j$ should be interpreted as effective interval-wise dark-energy-density parameters rather than as direct samples of a single globally tied function $X(z)$. The advantage of this localized construction is that the resulting bins are nearly uncorrelated, so the final constraints behave much like band powers in redshift. They are also deliberately conservative: by projecting out the absolute low-redshift transverse mode and by not enforcing a globally tied cosmological model in the baseline fit, we sacrifice some constraining power in exchange for a cleaner, more weakly model-dependent interpretation. Likewise, the analysis is not fully prior-free, because the local nuisance sector is regularized by broad external priors. Our aim here is therefore modest and explicit: to provide a low-redshift-agnostic BAO baseline, clarify what information is being removed and retained, and quantify the resulting interval-wise constraints from current DESI BAO measurements.

This paper is organized as follows. Section~\ref{sec:method} introduces the compressed BAO observables, the node-matched binning for $X(z)$, and the likelihood used in our baseline analysis. Section~\ref{sec:data} summarizes the DESI DR1 and DR2 BAO inputs. Section~\ref{sec:results} presents the interval-wise constraints and their correlation matrices. Section~\ref{sec:conclusion} discusses the interpretation, limitations, and future extensions of the method.

\section{Methodology}
\label{sec:method}

\subsection{Background expansion and BAO observables}
\label{sec:background}

We work in a spatially flat background and neglect radiation over the redshift range relevant to the BAO measurements used here. The dimensionless expansion history is
\begin{equation}
E(z) \equiv \frac{H(z)}{H_0},
\end{equation}
and we write
\begin{equation}
E^2(z) = \Omega_m (1+z)^3 + (1-\Omega_m) X(z),
\label{eq:E2_flat}
\end{equation}
where $\Omega_m$ is the present-day matter density parameter and $X(z)$ is the normalized dark-energy density evolution.

Because BAO distances are reported in units of the sound horizon at baryon drag, it is convenient to define the nuisance combination
\begin{equation}
S \equiv H_0\,\rd,
\label{eq:S_def}
\end{equation}
so that the overall distance scale appears as $c/S$. The anisotropic BAO observables are then
\begin{align}
\frac{\DHub(z)}{\rd} &= \frac{c}{S}\,\frac{1}{E(z)},
\label{eq:DH_over_rd}\\
\frac{\DM(z)}{\rd} &= \frac{c}{S}\,\int_{0}^{z} \frac{dz'}{E(z')}.
\label{eq:DM_over_rd}
\end{align}
Equation~(\ref{eq:DM_over_rd}) makes the key nonlocality explicit: each transverse-distance point contains the full integral from $z=0$ to $z$.

\subsection{Low-redshift-free compression}
\label{sec:compression}

Assume that anisotropic BAO measurements are available at a set of effective redshifts (``nodes'')
\begin{equation}
\{z_i\}_{i=1}^{N}, \qquad z_1 < z_2 < \cdots < z_N.
\end{equation}
The standard BAO summary vector is
\begin{equation}
\bm{x} = \left(
\frac{\DM(z_1)}{\rd},\frac{\DHub(z_1)}{\rd},
\frac{\DM(z_2)}{\rd},\frac{\DHub(z_2)}{\rd},
\ldots,
\frac{\DM(z_N)}{\rd},\frac{\DHub(z_N)}{\rd}
\right)^{\mathsf{T}},
\label{eq:xvec}
\end{equation}
with covariance $\bm{C}_x$.

To remove the additive low-redshift contribution below $z_1$, we replace the set $\{\DM(z_i)/\rd\}$ by adjacent increments,
\begin{equation}
\Delta\DM(z_j,z_{j+1}) \equiv \DM(z_{j+1}) - \DM(z_j), \qquad j=1,\ldots,N-1,
\label{eq:deltaDM}
\end{equation}
and define the compressed vector
\begin{equation}
\bm{y} = \left(
\frac{\DHub(z_1)}{\rd},\ldots,\frac{\DHub(z_N)}{\rd},
\frac{\Delta\DM(z_1,z_2)}{\rd},\ldots,\frac{\Delta\DM(z_{N-1},z_N)}{\rd}
\right)^{\mathsf{T}}.
\label{eq:yvec}
\end{equation}
The dimension of $\bm{y}$ is $2N-1$, so relative to the original $2N$-dimensional BAO summary we have intentionally removed one mode.

The mapping $\bm{y} = \bm{T}\bm{x}$ is linear. In components,
\begin{align}
y_i &= x_{2i}, \qquad i=1,\ldots,N,
\label{eq:map_DH}\\
y_{N+j} &= x_{2(j+1)-1} - x_{2j-1}, \qquad j=1,\ldots,N-1.
\label{eq:map_dDM}
\end{align}
Therefore the covariance propagates exactly as
\begin{equation}
\bm{C}_y = \bm{T}\,\bm{C}_x\,\bm{T}^{\mathsf{T}}.
\label{eq:Cy}
\end{equation}
This compression preserves the information relevant to reconstructing the expansion history above $z_1$ after projecting out the absolute transverse-distance offset associated with $z<z_1$.

\subsection{Node-matched binning and effective interval parameters}
\label{sec:binning}

We parameterize $X(z)$ as piecewise constant on the same redshift intervals defined by adjacent BAO nodes,
\begin{equation}
X(z) = X_j, \qquad z_j \le z < z_{j+1}, \qquad j=1,\ldots,N-1.
\label{eq:X_piecewise}
\end{equation}
This node-matched binning is chosen so that each interval parameter is constrained only by observables drawn from the same redshift range.

In addition to the $X_j$, our baseline analysis introduces local nuisance parameters $(\Omega_{m,j}, S_j)$ for each interval. In a globally consistent cosmology one would impose $\Omega_{m,j} = \Omega_m$ and $S_j = S$ for all $j$. We do not impose those equalities here, because our present goal is to localize the inference and isolate which combinations of the compressed BAO data constrain each redshift interval. The price of this flexibility is interpretational: with local nuisance parameters, the fitted $X_j$ are best understood as effective interval-wise descriptors of the expansion history rather than as direct samples of a single globally tied function $X(z)$.

For the $\DHub/\rd$ nodes we adopt a left-continuous assignment in which the node at $z_i$ is associated with the interval immediately below it,
\begin{equation}
b(i) =
\begin{cases}
1, & i=1, \\
i-1, & i=2,\ldots,N.
\end{cases}
\label{eq:binassign}
\end{equation}
With this convention, each interval contributes one local $\Delta\DM/\rd$ measurement across the interval and one $\DHub/\rd$ node at its upper edge.

\subsection{Model prediction, priors, and likelihood}
\label{sec:like}

The full parameter vector is
\begin{equation}
\theta = \{X_1,\ldots,X_{N-1},\,\Omega_{m,1},\ldots,\Omega_{m,N-1},\,S_1,\ldots,S_{N-1}\}.
\end{equation}
The model prediction corresponding to Equation~(\ref{eq:yvec}) is
\begin{align}
\mu_i &= \frac{c}{S_{b(i)}}\,\frac{1}{E\bigl(z_i;\Omega_{m,b(i)},X_{b(i)}\bigr)}, \qquad i=1,\ldots,N,
\label{eq:mu_DH}\\
\mu_{N+j} &= \frac{c}{S_j}\int_{z_j}^{z_{j+1}} \frac{dz}{E\bigl(z;\Omega_{m,j},X_j\bigr)}, \qquad j=1,\ldots,N-1.
\label{eq:mu_dDM}
\end{align}
Within each interval, $E(z)$ is evaluated using Equation~(\ref{eq:E2_flat}) with $(\Omega_m, X) \rightarrow (\Omega_{m,j}, X_j)$.

We assume a multivariate Gaussian likelihood for the compressed vector,
\begin{equation}
-2\ln \mathcal{L}(\theta) = \left(\bm{y} - \bm{\mu}(\theta)\right)^{\mathsf{T}} \bm{C}_y^{-1} \left(\bm{y} - \bm{\mu}(\theta)\right) + \mathrm{const}.
\label{eq:lnL}
\end{equation}
We impose the top-hat priors
\begin{align}
X_j &\sim \mathcal{U}(-2,4), \\
\Omega_{m,j} &\sim \mathcal{U}(0.28,0.35), \\
S_j &\sim \mathcal{U}(9.423\times 10^3,\,1.0393\times 10^4)\;\mathrm{km\,s^{-1}},
\end{align}
where the $\Omega_{m,j}$ and $S_j$ intervals correspond to approximately $5\sigma$ around the Planck 2018 $\Lambda$CDM best-fit values \citep{Planck2018}. To ensure a real expansion rate, we require
\begin{equation}
E^2(z;\Omega_{m,j},X_j) > 0
\end{equation}
within each interval; in practice it is sufficient to check the bin endpoints.

It is important to state explicitly that this baseline analysis is not prior-free. For $N$ BAO nodes, the compressed data vector contains $2N-1$ observables, whereas the local model above contains $3(N-1)$ parameters. The nuisance sector is therefore regularized primarily by the adopted priors on $\Omega_{m,j}$ and $S_j$. Throughout this paper, the quoted $X_j$ constraints should be interpreted as conditional on those priors rather than as fully prior-independent BAO determinations.

\subsection{Numerical implementation and scope}
\label{sec:numerics}

The interval integrals in Equation~(\ref{eq:mu_dDM}) are evaluated independently for each bin using fixed-order Gauss--Legendre quadrature. Posterior sampling is performed with the No-U-Turn Sampler, and we report posterior means, standard deviations, and the full covariance or correlation matrix of the interval parameters. A Julia implementation accompanies this work\footnote{The Julia code is available at \url{https://github.com/icosmology/BAO-DE-fit/}.}.

The present paper is intentionally focused on the compression scheme and its application to the DESI BAO summary measurements. We do not yet present a dedicated mock-based validation, a systematic comparison with a globally tied $(\Omega_m,S)$ model, or an exhaustive prior-sensitivity study. These tests are important next steps and are discussed again in Section~\ref{sec:conclusion}.

\section{DESI DR1 and DR2 BAO measurements}
\label{sec:data}

DESI is a 5000-fiber multi-object spectroscopic survey on the Mayall 4-m telescope at Kitt Peak National Observatory, designed to map the three-dimensional distribution of galaxies and quasars over a wide redshift range \citep{DESI_Survey,DESI_Instrument}. We use the public anisotropic BAO distance constraints released by the DESI Collaboration in DR1 and DR2 \citep{DESI_DR1_BAO,DESI_DR2_BAO}. For broader dark-energy interpretations of the DR2 cosmology results, see also the recent companion analyses by \citet{DESI_DR2_ExtendedDE} and \citet{DESI_DR2_DynamicalDE}. In each effective-redshift bin, the BAO likelihood is summarized by the pair $(\DM/\rd,\DHub/\rd)$ and its $2\times 2$ covariance, or equivalently by the two uncertainties and their correlation coefficient.

For DR1 our compilation contains $N=5$ nodes at $\{z_i\} = \{0.51,\,0.71,\,0.93,\,1.32,\,2.33\}$, while for DR2 it contains $N=6$ nodes at $\{z_i\} = \{0.51,\,0.706,\,0.934,\,1.321,\,1.484,\,2.33\}$. These nodes span the DESI galaxy and quasar BAO measurements together with the high-redshift Ly$\alpha$-forest BAO constraint \citep{DESI2024.III.KP4,DESI2024.IV.KP6,DESI.DR2.BAO.lya}.

For the published anisotropic BAO summaries used here, measurements from different effective-redshift bins are treated as uncorrelated, so the full input covariance is block diagonal. This is a property of the BAO analysis itself rather than an extra approximation introduced by our method. The same linear map in Equations~(\ref{eq:map_DH})--(\ref{eq:Cy}) would also propagate any more general covariance structure if needed.

\section{Results}
\label{sec:results}

\subsection{Effective interval-wise constraints on $X(z)$}
\label{sec:xconstraints}

\begin{figure}[!t]
\centering
\plotorplaceholder{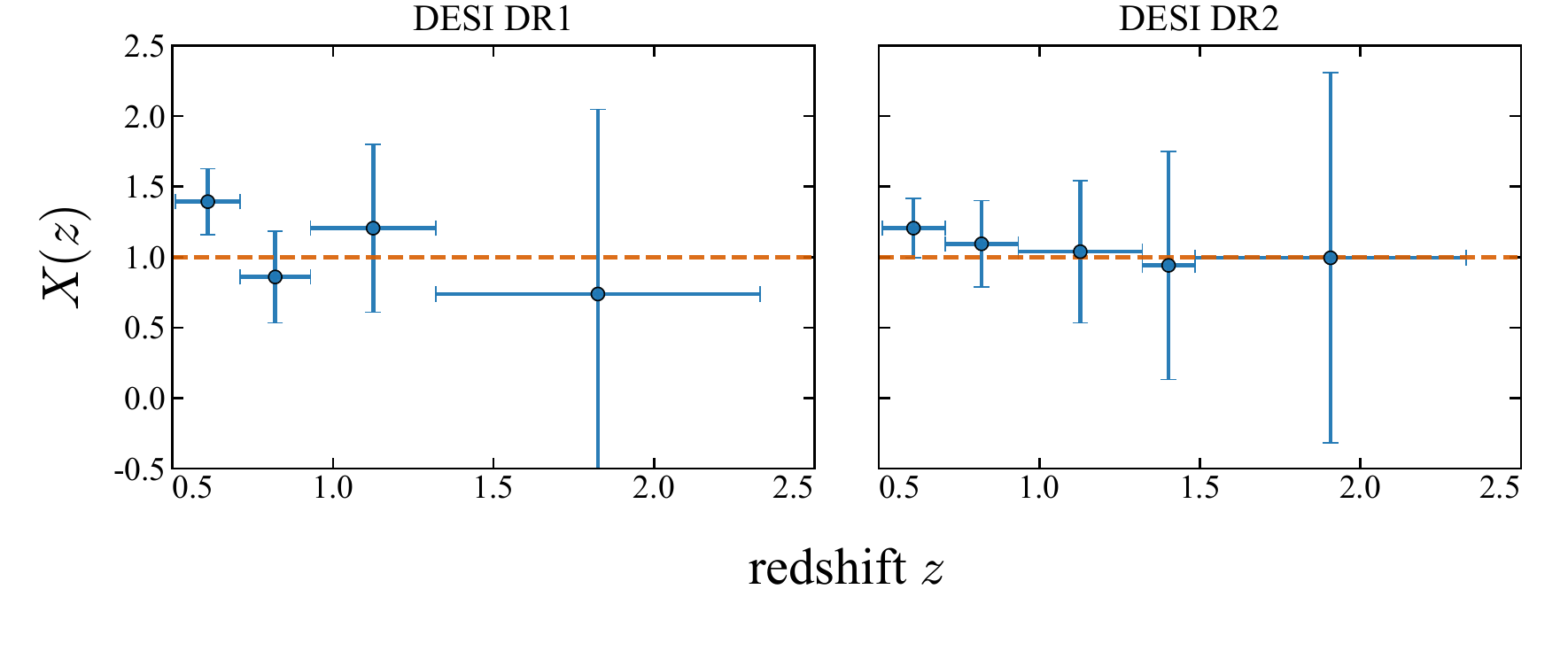}{0.92\textwidth}
\caption{Constraints on the effective interval-wise parameters $X_j$ inferred from DESI anisotropic BAO distances using the low-redshift-free compression $\{\DM(z_i)/\rd\} \rightarrow \{\Delta\DM(z_i,z_{i+1})/\rd\}$. The binning is matched to the BAO nodes, with $X(z)=X_j$ for $z_j \le z < z_{j+1}$ and plotted at $\bar z_j = (z_j + z_{j+1})/2$. The error bars are conditional on the priors described in Section~\ref{sec:like}.}
\label{fig:Xz}
\end{figure}

Figure~\ref{fig:Xz} summarizes the constraints on the interval parameters $X_j$ inferred from DESI DR1 and DR2 using the low-redshift-free compressed likelihood. Because the baseline model adopts local nuisance parameters in each interval, these $X_j$ should be interpreted as effective interval-wise constraints rather than as samples of a globally tied function $X(z)$. \rev{Here $\bar z_j \equiv (z_j + z_{j+1})/2$ denotes the arithmetic midpoint of the $j$th interval, used only to place each $X_j$ constraint on the horizontal axis.}

The constraints are strongest at low redshift and degrade rapidly toward higher redshift, as expected from the current BAO precision. For DR1 we find $X_1 = 1.39 \pm 0.23$ at $\bar z_1 \simeq 0.61$, corresponding to a mild $\sim 1.7\sigma$ upward fluctuation relative to $X=1$. For DR2 the lowest-redshift interval gives $X_1 = 1.21 \pm 0.21$ at $\bar z_1 \simeq 0.61$, consistent with $X=1$ at roughly the $1\sigma$ level. At higher redshift the typical uncertainty broadens from $\sigma(X) \simeq 0.2$--$0.3$ at $z \lesssim 0.9$ to $\sigma(X) \gtrsim 0.5$ for $z \gtrsim 1$, reaching order unity in the highest-redshift bins. All fitted intervals are consistent with $X=1$ within their $1\sigma$ uncertainties. Numerical values are listed in Table~\ref{tab:Xbins}.

\begin{table*}[!t]
\centering
\caption{Posterior means and $68\%$ credible intervals for the effective interval-wise parameters $X_j$ inferred from DESI DR1 and DR2. Each $X_j$ is defined for $z_j \le z < z_{j+1}$ and plotted at $\bar z_j = (z_j + z_{j+1})/2$. The quoted intervals are conditional on the priors described in Section~\ref{sec:like}.}
\label{tab:Xbins}
\begin{tabular}{lccccc}
\hline\hline
Data & Bin & $z_j$ & $z_{j+1}$ & $\bar z_j$ & $X_j$ \\
\hline
DR1 & 1 & 0.51 & 0.71 & 0.61  & $1.39\pm0.23$ \\
DR1 & 2 & 0.71 & 0.93 & 0.82  & $0.86\pm0.33$ \\
DR1 & 3 & 0.93 & 1.32 & 1.125 & $1.21\pm0.60$ \\
DR1 & 4 & 1.32 & 2.33 & 1.825 & $0.74\pm1.31$ \\
\hline
DR2 & 1 & 0.51  & 0.706 & 0.608 & $1.21\pm0.21$ \\
DR2 & 2 & 0.706 & 0.934 & 0.820 & $1.09\pm0.31$ \\
DR2 & 3 & 0.934 & 1.321 & 1.128 & $1.04\pm0.51$ \\
DR2 & 4 & 1.321 & 1.484 & 1.403 & $0.94\pm0.81$ \\
DR2 & 5 & 1.484 & 2.33  & 1.907 & $1.00\pm1.31$ \\
\hline\hline
\end{tabular}
\end{table*}

\subsection{Correlations, locality, and conservative band-power interpretation}
\label{sec:correlations}

\begin{figure}[!t]
\centering
\plotorplaceholder{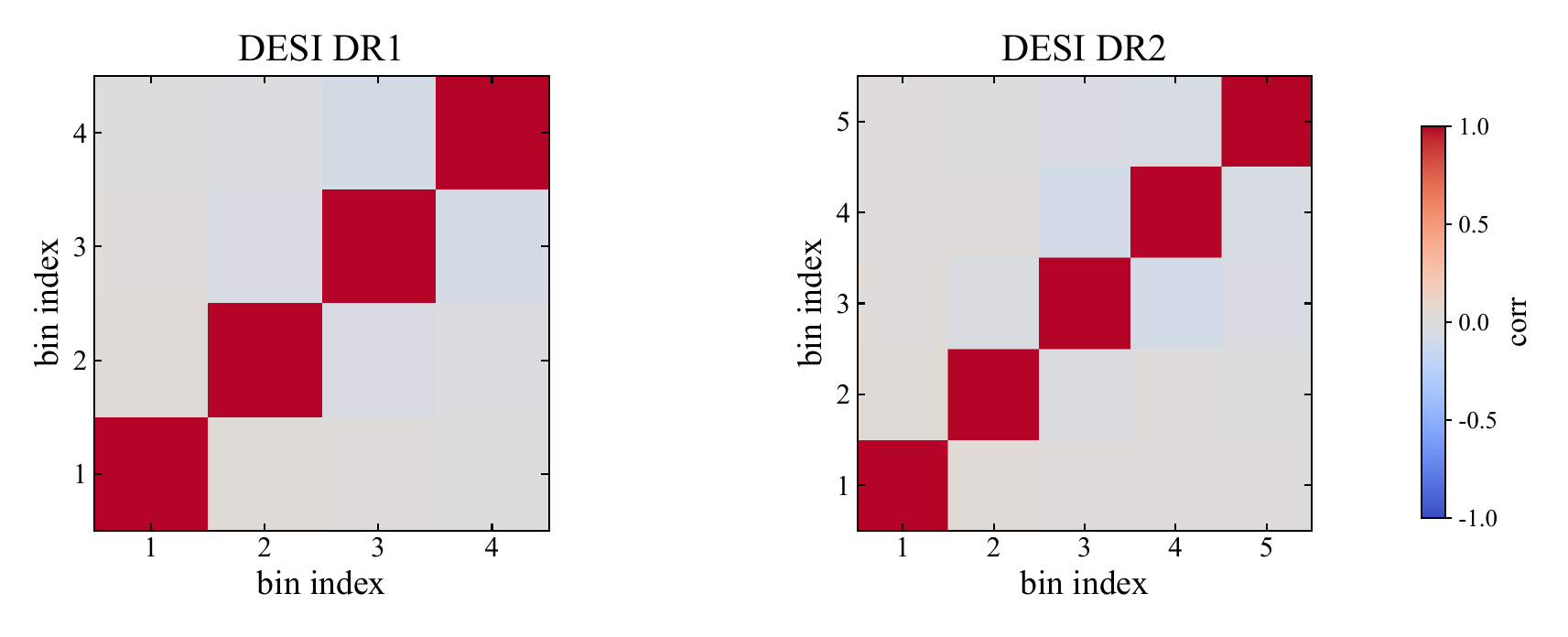}{0.98\textwidth}
\caption{Correlation matrices among the effective interval-wise parameters $X_j$ inferred from the baseline DESI DR1 (left) and DR2 (right) analyses. The weak off-diagonal structure reflects the combined effect of the low-redshift-free compression, the redshift locality of the BAO measurements, and the use of bin-local nuisance parameters.}
\label{fig:Corr}
\end{figure}

Figure~\ref{fig:Corr} shows the correlation matrices among the fitted $X_j$ parameters. In the baseline setup, the inferred correlations are weak: the maximum off-diagonal correlation is $|\rho| \lesssim 0.057$ for DR1 and $|\rho| \lesssim 0.063$ for DR2. This near-diagonal structure is one of the main practical advantages of the method. It makes the interval-wise results easy to interpret and to combine with external information, and it motivates viewing the $X_j$ as conservative band-power-like estimates of dark-energy evolution in redshift space.

The term ``conservative'' is important here. Each interval is constrained primarily by BAO information drawn from the same redshift range, while the absolute low-redshift transverse-distance mode has been projected out and stronger global assumptions have not been imposed in the baseline fit. The price is larger error bars than one would obtain in a more tightly tied cosmological reconstruction, but the payoff is a cleaner and more local interpretation of what each bin measures.

These weak correlations should not be attributed to the compression alone. They reflect the combined effect of the low-redshift-free $\Delta\DM$ construction, the intrinsic redshift locality of the BAO summary itself, and the choice to fit bin-local nuisance parameters. In other words, the near-diagonal $X_j$ covariance is not an artificial consequence of an ad hoc covariance truncation; it is the natural outcome of applying this localized reconstruction to BAO measurements that are already essentially uncorrelated across redshift bins.

\subsection{Interpretation of the nuisance sector}
\label{sec:nuisance}

The local nuisance parameters $(\Omega_{m,j},S_j)$ are introduced to regularize and localize the inference; they should not be interpreted as independent measurements of the matter density or the sound-horizon scale in each redshift interval. Given the dimensionality of the local model, their posteriors remain substantially influenced by the adopted priors. The same caveat carries over, to a lesser degree, to the inferred $X_j$ constraints.

For this reason, we regard Table~\ref{tab:Xbins} and Figure~\ref{fig:Corr} as a transparent baseline conditioned on the priors stated in Section~\ref{sec:like}. Precisely because the bins are nearly uncorrelated and intentionally conservative, these results are useful as a band-power-like summary that can be compared directly with more model-driven reconstructions. A useful next step will be to compare them with two complementary analyses: first, a physically tied fit with common $(\Omega_m,S)$ across all bins; and second, a broader-prior study that quantifies how robust the $X_j$ constraints are to relaxing the external regularization of the nuisance sector.

\section{Discussion and conclusion}
\label{sec:conclusion}

We have presented a low-redshift-agnostic compression of anisotropic BAO distances for reconstructing binned dark-energy-density evolution above the lowest BAO node \rev{$z_1$}. The core idea is simple: replace the standard transverse-distance set $\{\DM(z_i)/\rd\}$ by adjacent increments $\{\Delta\DM(z_i,z_{i+1})/\rd\}$ while retaining the radial nodes $\{\DHub(z_i)/\rd\}$. Because this transformation is linear, the covariance matrix propagates exactly. Relative to the original BAO summary vector, the method intentionally removes one absolute transverse-distance mode---the additive contribution from $z<z_1$---and keeps the remaining information relevant to reconstructing the expansion history above that redshift.

Applied to DESI DR1 and DR2, the method yields effective interval-wise constraints on $X(z)$ that are strongest at low redshift and weaken substantially toward $z\gtrsim 1$. All fitted intervals are consistent with $X=1$ within current uncertainties. In the baseline implementation adopted here, the inferred interval parameters are only weakly correlated, which is a central advantage of the method. The resulting $X_j$ therefore behave like conservative band powers in redshift: they are easy to summarize, straightforward to compare across bins, and relatively clean to combine with external information. In this sense, the present analysis is complementary to recent DESI DR2 dark-energy studies that use broader data combinations and more model-driven reconstructions \citep{DESI_DR2_ExtendedDE,DESI_DR2_DynamicalDE}.

The analysis nevertheless has important limitations. First, the fitted $X_j$ are effective interval parameters because each redshift bin is assigned its own nuisance parameters $(\Omega_{m,j},S_j)$ rather than sharing a single globally consistent cosmology. Second, the nuisance sector is regularized by broad external priors, so the quoted $X_j$ intervals are not fully prior-free BAO determinations. Third, although the BAO summary used here is naturally block diagonal across redshift bins, we do not yet present a dedicated mock-based validation of the reconstruction pipeline. These caveats do not invalidate the baseline results, but they do define their proper interpretation.

The most important next steps are therefore clear. A comparison between local and globally tied nuisance-parameter models would help isolate how much additional localization is produced by the compression and bin-local fit beyond the intrinsic redshift locality of the BAO measurements themselves. A broader prior-sensitivity analysis would quantify the robustness of the inferred $X_j$ values. Mock tests would provide an end-to-end check of bias and covariance propagation. More generally, the present framework can serve as a useful bridge between raw anisotropic BAO distance measurements and more physically constrained reconstructions of dark-energy evolution, while already offering a conservative, nearly uncorrelated band-power summary of the current BAO information.

\begin{acknowledgements}
G.-B. Zhao is supported by National Natural Science Foundation of China (NSFC) grants 12525301, 12273048, and 12422301, and by the Youth Innovation Promotion Association of the Chinese Academy of Sciences.
\end{acknowledgements}

\label{lastpage}

\bibliographystyle{raa}
\bibliography{refs}
\end{document}